\documentstyle[aps,pre,epsf]{revtex}
\begin{document}
\title{Statistics of two-particle dispersion in two-dimensional turbulence}
\author{G. Boffetta}
\address{Dipartimento di Fisica Generale and INFM, Universit\`a di Torino \\
via Pietro Giuria 1, 10125 Torino, Italy}
\author{I.~M. Sokolov}
\address{Theoretische Polymerphysik, Universit$\ddot{a}$t Freiburg, \\
Hermann-Herder Sta$\ss$e 3 \\
D-79104 Freiburg i.Br., Germany}
\date{\today}
\maketitle

\begin{abstract}
We investigate Lagrangian relative dispersion in direct numerical
simulation of two-dimensional inverse cascade turbulence. 
The analysis is performed by using
both standard fixed time statistics and an exit time approach. Our results
are in good agreement with the original Richardson's description in terms of
diffusion equation. The deviations are only of the quantitative nature. These
deviations, and the observed strong sensitivity to the finite size effects, 
are associated with the long-range-correlated nature of the particles' 
relative motion.
The correlation, or persistence, parameter is measured by means of
a Lagrangian ``turning point'' statistics.
\end{abstract}



\section{Introduction}

\label{sec:1} Understanding the statistics of particle pairs dispersion in
turbulent velocity fields is of great interest for both theoretical and
practical implications. At variance with single particle dispersion which
depends mainly on the large scale, energy containing eddies, pair dispersion
is driven (at least at intermediate times) by velocity fluctuations at
scales comparable with the pair separation. Since these small scale
fluctuations have universal characteristics, independent from the details of
the large scale flow, relative dispersion in fully developed turbulence is
expected to show universal behavior \cite{Batchelor52,MY75}. From an
applicative point of view, a deep comprehension of relative dispersion
mechanisms is of fundamental importance for a correct modelization of small
scale diffusion and mixing properties.

Since the pioneering work by Richardson \cite{Richardson26}, many efforts
have been done to confirm experimentally or numerically his law 
\cite{MY75,ZB94,EM96,FV98,JPT99,MOA99,BCCV99,BC00}. Nevertheless, the main
obstacle to a deep investigation of relative dispersion in turbulence
remains the lack of sufficient statistics due to technical difficulties in
laboratory experiments and to the moderate inertial range reached in direct
numerical simulations.

In this Paper we present a detailed investigation of the statistics of
relative dispersion from extensive direct numerical simulations of particle
pairs in two-dimensional Navier-Stokes turbulence. We will see that the main
ingredient of the original Richardson description, i.e. Richardson diffusion
equation, is sufficient for rough description of relative dispersion in this
flow. Nevertheless, our simulations show that two-particle statistics is
rather sensible to finite size effects. This asks for a different analysis
based on doubling time statistics which has recently introduced for the
analysis of relative dispersion \cite{ABCCV97}. Comparison of numerical
results with ones based on the Richardson's equation show that the last
delivers a qualitatively good description of the
doubling-time-distributions. The quantitative deviations found are
attributed to the fact that the dispersion process is not purely diffusive
and is influenced by ballistic (persistent) motion.

The article is organized as follows: In Section \ref{sec:2} we discuss the
Richardson's approach to the two-particle dispersion, in Section~\ref{sec:3}
the fixed-scale properties of dispersion process (such as doubling-time
statistics) are considered. The numerical approach and the results of
simulations are discussed in Section~\ref{sec:4}. Section~\ref{sec:5} is
devoted to conclusions. The mathematical details of calculations of
doubling-time statistics for the Richardson's case are
given in Appendices \ref{app:1} and \ref{app:2}.


\section{Statistics of relative dispersion}
\label{sec:2}
Relative dispersion in turbulence is often
phenomenologically described in terms of a diffusion equation for the
probability density function of pair separation $p({\bf r},t)$ 
\begin{equation}
\frac{\partial p({\bf r},t)}{\partial t}=\frac{\partial }{\partial r_{j}}%
\left( K({\bf r},t)\frac{\partial p({\bf r},t)}{\partial r_{j}}\right) 
\label{eq:2.1}
\end{equation}
with a space and time dependent diffusion coefficient $K({\bf r},t)$ \cite
{MY75}. The original Richardson proposal obtained from experimental data in
the atmosphere corresponds to $K(r,t)=K(r)=k_{0}\varepsilon ^{1/3}r^{4/3}$ 
\cite{Richardson26}, where $\varepsilon $ has the dimension of energy
dissipation (see below) and $k_{0}$ is an adimensional constant. In the $d$%
-dimensional isotropic case, this diffusion equation takes the form 
\begin{equation}
\frac{\partial p({\bf r},t)}{\partial t}={\frac{1}{r^{d-1}}}
{\frac{\partial}{\partial r}}r^{d-1}K(r)
{\frac{\partial p({\bf r},t)}{\partial r}}
\label{eq:2.2}
\end{equation}
Its solution leads to the well known non-Gaussian distribution 
\begin{equation}
p({\bf r},t)={\frac{A}{(k_{0}t)^{3}\varepsilon }}exp
\left( -{\frac{9r^{2/3}}{4k_{0}\varepsilon ^{1/3}t}}\right)   
\label{eq:2.3}
\end{equation}
where $A$ is a normalizing factor. The growth of pair separation is
described in terms of a single exponent as 
\begin{equation}
R^{2n}(t)\equiv \langle r^{2n}(t)\rangle =C_{2n}\varepsilon ^{n}t^{3n}.
\label{eq:2.4}
\end{equation}
The so called Richardson constant in (\ref{eq:2.4}) is 
$g=C_{2}={\frac{1280}{243}}k_{0}^{3}$.

The Richardson's conjecture was formulated based on the
scaling nature of the diffusion coefficient and in analogy with
diffusion. No information about the nature of the pdf was available by that
time. Any chose of the form 
$K(r,t)\simeq r^{4/3-\alpha }\langle r^{2}(t)\rangle ^{\alpha /2}$ 
(i.e. $K(r,t)\simeq r^{4/3-\alpha }t^{3\alpha /2}$) 
would give the same scaling law $R^{2}\propto t^{3}$ but with different pdf's
(see \cite{Batchelor52,MY75,Okubo62,K66}).

The possibility to describe the dispersion process by means of a diffusion
equation is based on essentially two important physical assumptions which
can be verified a posteriori. The first one is that the dispersion process
is self-similar in time, which is probably true in non intermittent velocity
field \cite{BCCV99}, the second one is that the velocity field is short
correlated in time \cite{S99}. 
In the limit of velocity field $\delta$-correlated in time the diffusion
equation (\ref{eq:2.1}) becomes exact \cite{K68,GV00}.
As we proceed to show, the Richardson's
conjecture (\ref{eq:2.2}), which is exact under small values of the
persistence parameter of the flow, {\it vide infra} and \cite{S99,SKB00},
still delivers a qualitatively good approximation for realistic 2d turbulent
flows, whose persistence parameter of the order of 1.

Richardson scaling in turbulence is a consequence of Kolmogorov scaling for
the velocity differences \cite{MY75}. Under Kolmogorov scaling, the mean
square relative velocity and the correlation time in the inertial range are
given by 
\begin{equation}
\langle \delta v(r)^{2}\rangle =v_{0}^{2} \left( {\frac{r}{r_{0}}}\right)
^{2/3} \simeq \varepsilon^{2/3} r^{2/3}  \label{eq:2.5}
\end{equation}
and 
\begin{equation}
\tau (r)=\tau _{0}\left( {\frac{r}{r_{0}}}\right)^{2/3} \simeq
\varepsilon^{-1/3} r^{2/3} \, ,  \label{eq:2.6}
\end{equation}
where $r_{0}$, $\tau _{0}$ and $v_{0}$ are some (large scale) characteristic
length, time and velocity scale and $\varepsilon \simeq v_{0}^2/\tau_{0}$ is
the energy flux in the inertial range. The value of the dimensionless
combination $Ps=\frac{v_{0}\tau _{0}}{r_{0}}$ remains however unspecified by
scaling considerations. It is referred to as a persistence parameter of the
flow and plays a central role in describing single particle diffusion and
pair separation \cite{S99,SKB00}.

The persistence parameter gives the ratio of the velocity correlation time
to the Lagrangian characteristic time. In order to see how $Ps$ influences
Lagrangian dispersion, let us consider the following simple model, which has
been used as a basis for building a stochastic model of turbulent dispersion 
\cite{S99}. We take that the magnitude of the separation velocity is a
function of $r$ only so that $\delta v(r)=v_{0}(r/r_{0})^{1/3}$. The
temporal changes of the flow can be accounted for by letting the particle
change its velocity direction from time to time, while keeping the
velocity's magnitude constant. The probability that the relative velocity of
particle separation changes its direction during time interval $dt$ can be
assumed to be $dp=dt/\tau (r)$. The growth of the magnitude of the
interparticle separation ${\bf r}(t)$ in $dt$ is $dr\simeq \delta v(r)dt$,
thus the probability to change the direction of velocity within $dr$ is,
using (\ref{eq:2.5}-\ref{eq:2.6}) 
\begin{equation}
dp=p(r)dr={\frac{dr}{\delta v(r)\tau (r)}}=
{\frac{r_{0}}{v_{0}\tau _{0}}}{ \frac{dr}{r}}={\frac{1}{Ps}}{\frac{dr}{r}.}  
\label{eq:2.7}
\end{equation}
The distribution of the position of turning points in the separation follows
from (\ref{eq:2.7}) \cite{SKB00}. The conditional probability density to
find a next turning point at $r_{2}$ provided a previous one was at 
$r_{1}<r_{2}$ is given by 
\begin{equation}
\Psi (r_{2}|r_{1})={\frac{1}{Ps\,r_{1}}}
\left( {\frac{r_{2}}{r_{1}}}\right)^{-1/Ps-1} \, .
\label{eq:2.8}
\end{equation}
Note that the dependence of $\Psi (r_{2}|r_{1})$ only on
the relative positions of the turning points, i.e. on $r_{2}/r_{1}$, is a
clear consequence of scaling.

The tail of $\Psi (r_{2}|r_{1})$ decides about the
existence of the second moment of this distribution, i.e. on the fact
whether the corresponding motion is short- or long-range correlated in
space. Depending on the persistence parameter $Ps$, the dispersion can be
either diffusive ($Ps\ll 1$) or ballistic ($Ps\gg 1$) in nature. Note that
the power-law tail of the distributions make the problem extremely sensitive
to the finite-size effects, especially for large $Ps$, when the weights of
ballistic events (L\'{e}vy-walks \cite{SWK87}) is considerable.

We note that the value of $Ps$ is not a free parameter, but is fixed for a
given physical situation. The scaling nature of turbulence supposes that
this parameters is a constant, depending only on general properties of the
flow, e.g. on its 2d or 3d nature (in this last case also the overall
geometry of the flow can be of importance). On the other hand, since the
nature of dispersion process depends crucially on the value of $Ps$, the
only way for getting quantitative information about the dispersion is
through direct numerical simulations.
The strong finite-size effects in relative dispersion simulations
require the introduction of quantities which are less sensitive 
to finite resolution.


\section{Exit time statistics}

\label{sec:3} All the results discussed in the previous Section can be
observed only in high Reynolds number flows in which the inertial range,
where the scaling laws (\ref{eq:2.5},\ref{eq:2.6}) hold, is sufficiently
wide. The needs for large Reynolds is particularly severe in the case of
Richardson dispersion, as a consequence of the long tails in (\ref{eq:2.3}).
Moreover, the observation of time scaling laws as (\ref{eq:2.4}) requires
sufficiently long times in order to forget the initial separation $r(0)$ 
\cite{MY75}.

For these reasons, the observation of Richardson scaling (i.e. (\ref{eq:2.3}%
) or (\ref{eq:2.4})) is very difficult in direct numerical simulations where
the Reynolds number is limited by the resolution. The same kind of
limitations arise in laboratory experiments, as a consequence of the
necessity to follow the Lagrangian trajectories which limits again the
Reynolds number \cite{JPT99,MOA99}.

To partially overcome these difficulties, an alternative approach based on 
{\it doubling time} (or exit time) statistics has been recently proposed 
\cite{BCCV99,ABCCV97}. Given a set of thresholds $R_{n}=\rho ^{n}R(0)$
within the inertial range, one computes the ``doubling time'' 
$T_{\rho}(R_{n})$ defined as the time it takes for the particle 
pair separation to
grow from threshold $R_{n}$ to the next one $R_{n+1}$. Averages are then
performed over many dispersion experiments, i.e., particle pairs, to get the
mean doubling time $\langle T_{\rho }(R)\rangle $. The outstanding advantage
of this kind of averaging at fixed scale separation, as opposite to a fixed
time, is that it removes crossover effects since all sampled particle pairs
belong to the same scales.

The problem of doubling time statistics is a first-passage problem for the
corresponding transport process. For the Richardson case, in 2d it is given
by the solution of the Richardson's diffusion equation, Eqs. (\ref{eq:2.2}),
with initial condition $p({\bf r},0)=\delta (r-R/\rho )/2\pi $ and absorbing
boundary at $r=R$ (so that $p(R,t)=0$). The pdf of doubling time can be
obtained as the time derivative of the probability that the particle is
still within the threshold 
\begin{equation}
p_{D}(t)=-{\frac{d}{dt}}\int_{|{\bf r}|<R}p({\bf r},t)d{\bf r}\,.
\label{eq:3.2}
\end{equation}
Using (\ref{eq:2.2}) one obtains 
\begin{equation}
p_{D}(t)=-2\pi \varepsilon ^{1/3}k_{0}R^{4/3}\left. 
{\frac{\partial p({\bf r},t)}{\partial r}}\right| _{r=R}.
\label{eq:3.3}
\end{equation}
The solution using the eigenfunction decomposition is given in Appendix~\ref
{app:1} and shows that the long-time asymptotic of $p_{D}(t)$ is
exponential: 
\begin{equation}
p_{D}(t)\simeq \exp (-\kappa k_{0}\varepsilon ^{1/3}R^{-2/3}t)
\label{eq:3.3b}
\end{equation}
where $\kappa \approx 2.93$ is a number factor. This exponential nature of
the tail of $p_{D}(t)$-distribution will be confirmed by direct simulations
in Section \ref{sec:4}.

Note that the combination $\varepsilon^{-1/3} R^{2/3}$ has a dimension of
time and is proportional to the average doubling time $\langle T_{\rho}(R)
\rangle$. This time can be obtained by a simple argument reported in the
Appendix~\ref{app:2}. In the $2$-dimensional case one obtains 
\begin{equation}
\langle T_{\rho }(R)\rangle ={\frac{3}{4}}{\frac{\rho ^{2/3}-1}{\varepsilon
^{1/3}\rho ^{2/3}}}\,\frac{R^{2/3}}{k_{0}}  \label{eq:3.4}
\end{equation}

Prediction (\ref{eq:3.4}) contains the parameter $k_{0}$ which is, as shown
in Section~\ref{sec:2}, is equivalent to the Richardson constant $g$. As a
consequence, the computation of average doubling time can be used for an
alternative (and more robust, as we will see) estimation of the Richardson
constant. It is convenient to rewrite the doubling time pdf (\ref{eq:3.3b})
in terms of the average doubling time $\langle T_{\rho }(R)\rangle $. Making
use of (\ref{eq:3.4}) one obtains in 2d 
\begin{equation}
p_{D}(t)\simeq \exp (-0.252 \, t/\langle T(R)\rangle )\,.
\label{eq:3.5}
\end{equation}
which is a parameterless, universal function.


\section{Direct numerical simulations}

\label{sec:4} Pair dispersion statistics has been investigated by
high-resolution direct numerical simulations of the inverse energy cascade
in two-dimensional turbulence \cite{KM80}. There are several reasons for
considering 2D turbulence. First of all, the dimensionality of the problem
makes feasible direct numerical simulations at high Reynolds numbers.
Moreover, the observed absence of intermittency \cite{BCV00} makes the 2D
inverse energy cascade an ideal framework for the study of Richardson
scaling in Kolmogorov turbulence.

The 2D Navier-Stokes equation for the vorticity 
$\omega=\nabla \times {\bf v} = -\Delta \psi$ is: 
\begin{equation}
\partial_t\omega+J\left(\omega,\psi\right)=\nu\Delta\omega 
-\alpha\omega + \phi,
\label{eq:4.1}
\end{equation}
where $\psi$ is the stream function and $J$ denotes the Jacobian. The
friction linear term $-\alpha \omega$ extracts energy from the system to
avoid Bose-Einstein condensation at the gravest modes \cite{SY93}. The
forcing is active only on a typical small scale $l_f$ and is $\delta$%
-correlated in time to ensure the control of the energy injection rate. The
viscous term has the role of removing enstrophy at scales smaller than $l_f$
and, as customary, it is numerically more convenient to substitute it by a
hyperviscous term (of order eight in our simulations). Numerical integration
of (\ref{eq:4.1}) is performed by a standard pseudospectral method on a
doubly periodic square domain at resolution $N=1024$ or $N=2048$. All the
results presented are obtained in conditions of stationary turbulence.

\begin{figure}[tbp]
\epsfbox{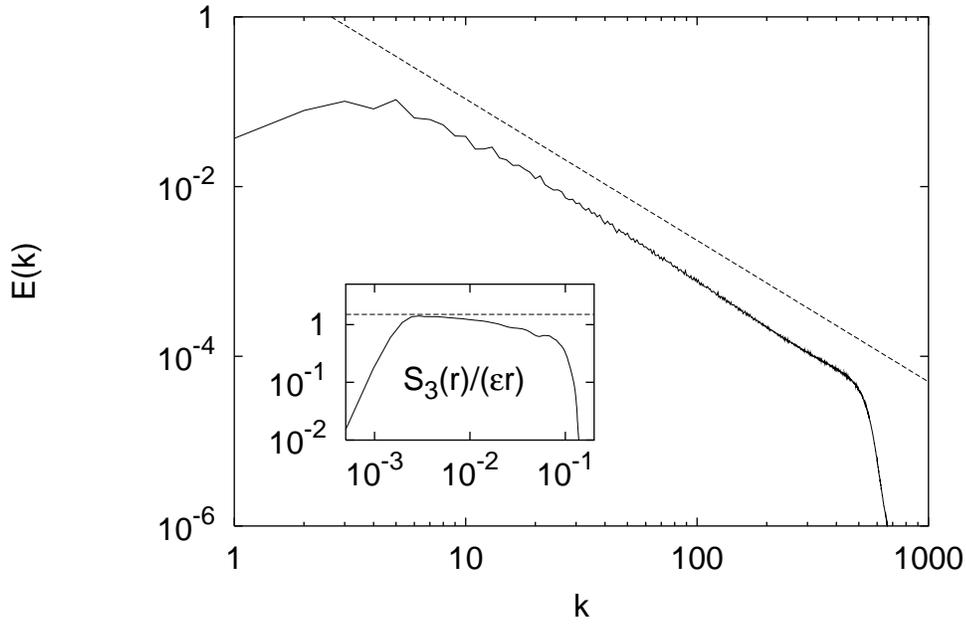}
\caption{ Energy spectrum of the inverse cascade simulations $E(k)$. The
dashed line has the Kolmogorov slope $-5/3$. In the inset it is shown the
compensated third order longitudinal structure function $S_{3}(r)/(%
\varepsilon r)$. The dashed line is the prediction $3/2$. }
\label{fig1}
\end{figure}

In Figure~\ref{fig1} we plot the energy spectrum, which displays
Kolmogorov scaling $E(k)=C \epsilon^{2/3} k^{-5/3}$ over about two decades
with Kolmogorov constant $C\simeq 6.0$. In the inset we plot the third order
structure function $S_{3}(r) = \langle \delta v(r)^3 \rangle$ compensated
with the theoretical prediction $S_3(r)=3/2 \varepsilon r$. The observation
of the plateau confirms the existence of an inverse energy cascade and
indicates the extension of the inertial range. Previous numerical
investigation has shown that velocity differences statistics in the inverse
cascade is not affected by intermittency corrections \cite{BCV00}. 
In this case we may expect the Lagrangian statistics to be self-similar 
with Richardson scaling \cite{BCCV99}.

Lagrangian statistics is obtained by integrating the trajectories of $64000$
particle pairs in the turbulent velocity field, initially uniformly
distributed with constant separation $R(0)$.

The Lagrangian data reported below are in dimensionless units in which
separations are rescaled with the box size $L=2\pi $ and time with the large
scale time $T_{0}=(L^{2}/\varepsilon )^{1/3}$.


\subsection{Relative dispersion analysis}
\label{sec:4.1} 
\begin{figure}[tbp]
\epsfbox{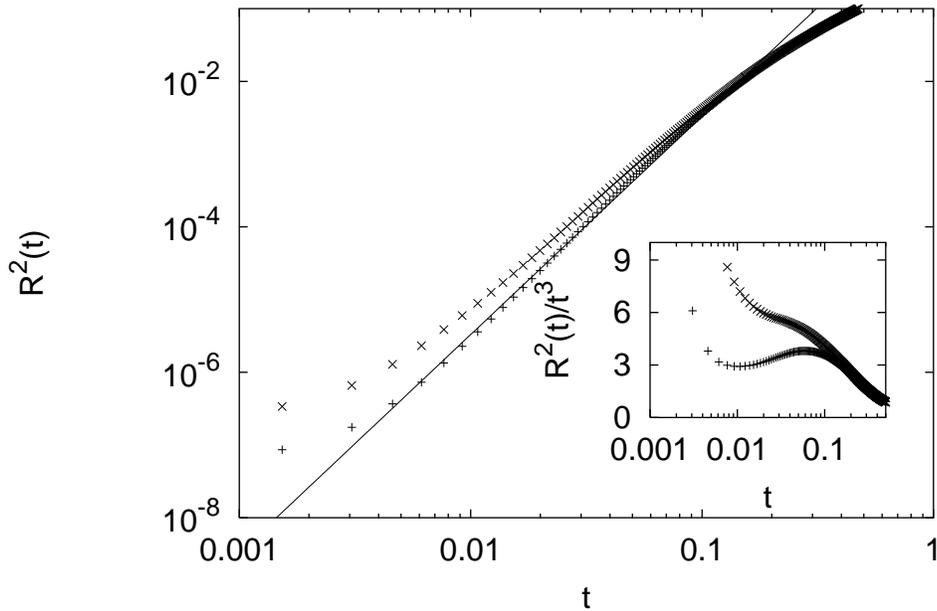}
\caption{ Relative dispersion $R^2(t)$ with $R(0)=\delta x/2$ ($+$) and $%
R(0)=\delta x$ ($\times$). In the inset the compensated plot $%
R^2(t)/(\varepsilon t^3)$. }
\label{fig2}
\end{figure}

In Figure~\ref{fig2} we plot the relative dispersion 
$R^{2}(t)$ for two different initial separation, 
$R(0)=\delta x/2$ and $R(0)=\delta x$ (where $\delta x=2\pi /N$ is 
the grid mesh). The Richardson $t^{3}$ law (\ref{eq:2.4}) is observed 
in a limited time interval, especially
for the larger $R(0)$ run. It is remarkable that the relative separation law
displays such a strong dependence on the initial conditions even in our high
resolution runs.

This dependence makes the determination of the Richardson constant
particularly difficult. In Figure~\ref{fig2} we also show the
compensated plot $R^{2}(t)/t^{3}$ which, in the adimensional units, should
directly give the constant $g$. It is clear that a precise determination of 
$g$ is impossible; even the Richardson scaling (\ref{eq:2.4}), when looked in
a compensated plot, is rather poor. Figure~\ref{fig2} suggests that
starting with an intermediate initial separation would give a wider scaling
range. Of course, one would like to avoid this ``fine tuning'', which is
probably impossible to implement in the case of experimental data.

\begin{figure}[tbp]
\epsfbox{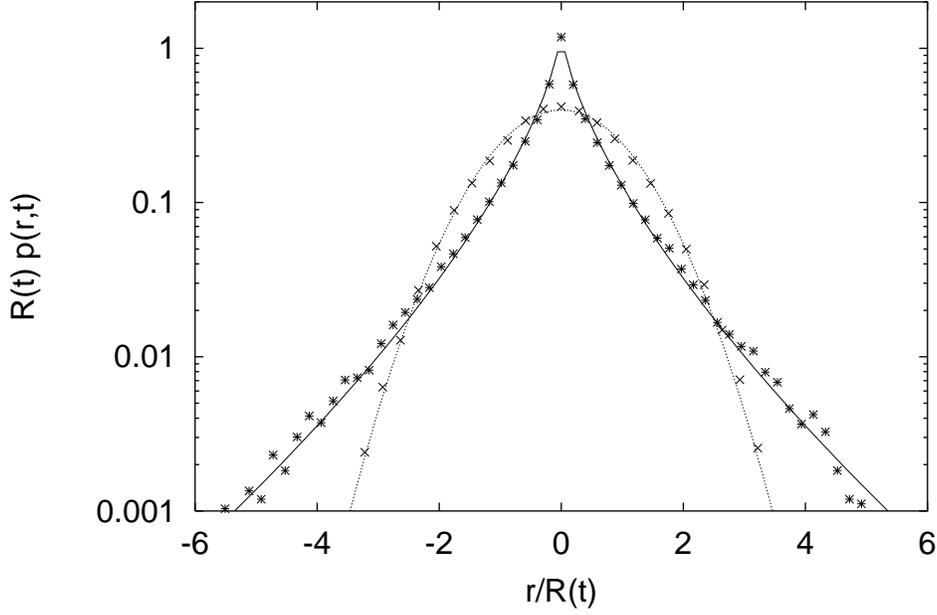}
\caption{ Probability distribution function of relative separations at times 
$t=0.031$ ($\ast$) and $t=0.77$ ($\times$) rescaled with $R(t)=\langle
r^{2}(t) \rangle^{1/2}$. The continuous line is the Richardson prediction (%
\ref{eq:2.3}), the dashed line is the Gaussian distribution. }
\label{fig3}
\end{figure}

\begin{figure}[tbp]
\epsfbox{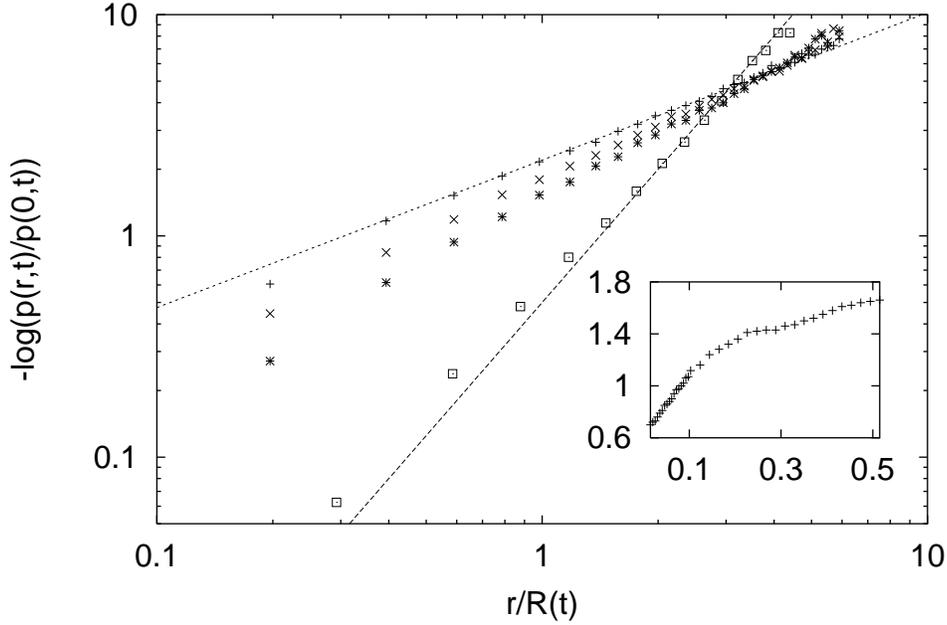}
\caption{ Right tail of $-\log(p(r,t)/p(0,t))$ at times $t=0.015$ ($+$), $%
t=0.041$ ($\times$), $t=0.067$ ($\ast$) and $t=0.77$ ($\Box$) in log-log
plot. The two lines represent the Richardson slope $2/3$ and the Gaussian
slope $2$. The inset shows the exponent of the right tail of the pdf as a
function of time. }
\label{fig4}
\end{figure}

The probability distribution function of pair separations is plotted in
Figure~\ref{fig3} for the $R(0)=\delta x/2$ run. At short time 
$t=0.015$, in the beginning of the $t^{3}$ range in 
Figure~\ref{fig2}, we found
that the Richardson pdf (\ref{eq:2.3}) fits pretty well our data, although
some deviations can be detected. Of course, at time comparable with 
the integral
time $t=0.77$, particle separations are of the order of the integral scale
and we observe Gaussian distribution. The crossover between these two
regimes is extremely broad: deviations from Richardson pdf are clearly seen
already for the times well within the Richardson's $t^{3}$ range. To observe
better this transition, in Figure~\ref{fig4} we plot, in log-log plot,
the right tail of $-\ln (p(r,t)/p(0,t))$. We see that the tail slope can be
fitted with an exponent $\alpha $ which change continuously in time, from 
$2/3$ to the Gaussian value $2$ (see the inset of Figure~\ref{fig4}).
Thus self similarity, if it exists, is reduced to the very short time at the
beginning of dispersion. Moreover, the scaling region is strongly affected
by the choice of initial separation, as shown in Figure~\ref{fig2}.

\subsection{Doubling time data}
\label{sec:4.2} 
\begin{figure}[tbp]
\epsfbox{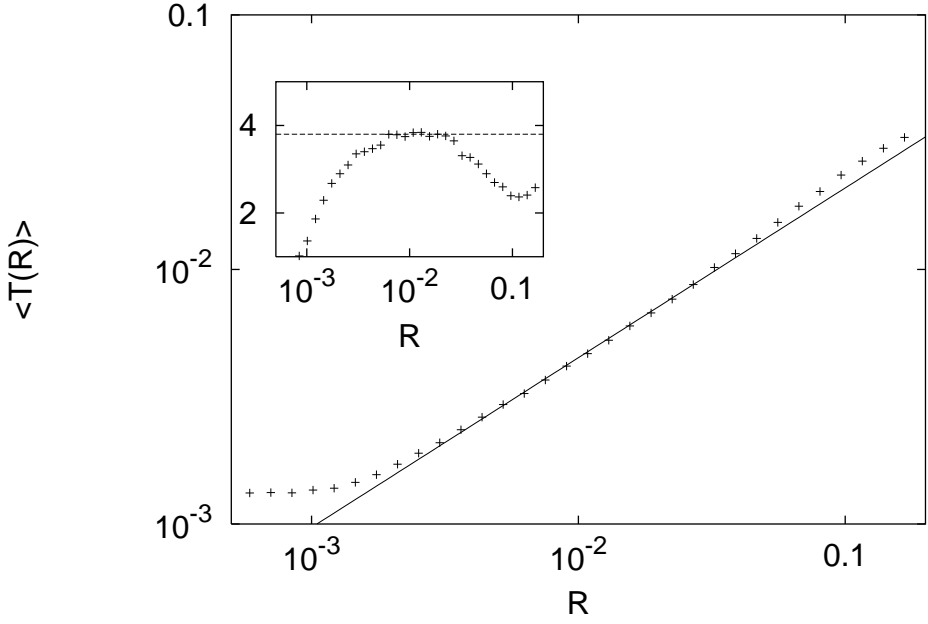}
\caption{ Mean doubling time $\langle T(R) \rangle$ as function of the
separation $R$. The ratio is $r=1.2$ and the average is obtained over about 
$5 \times 10^{5}$ events. The line represent the dimensional scaling $R^{2/3}$.
In the inset the compensated plot gives the value of Richardson constant 
$g \simeq 3.8$, as explained in the text. }
\label{fig5}
\end{figure}

From the same Lagrangian trajectories discussed in the previous
Section, we have computed the doubling time statistics. The average doubling
time as function of the scale is plotted in Figure~\ref{fig5}. The line
represent the dimensional prediction $\langle T(R)\rangle \simeq R^{2/3}$.
In the inset we plot the quantity 
${\frac{20}{9}}{\frac{(\rho ^{2/3}-1)^{3}}{\rho ^{2}}}
{\frac{R^{2}}{\varepsilon \langle T\rangle ^{3}}}$ which, from 
(\ref{eq:2.4}) and (\ref{eq:3.4}) gives the value of the Richardson 
constant $g\simeq 3.8$, which is compatible with the estimation from 
Figure~\ref{fig2}. It is interesting to compare our result with previous
estimations of $g$. The only experimental estimation of $g$ for 2D inverse
cascade \cite{JPT99} gives a value about $7$ times smaller (but the Reynolds
number in the experiment is even smaller than in present simulations). Other
estimations of $g$ are based on kinematic simulations with synthetic flows.
In all these cases \cite{EM96,FV98,FHMP92} the reported values are even
smaller. On the other hand, our numerical finding is not very far from the
prediction of turbulence closure theory \cite{K66,LL81}.

From Figure~\ref{fig5} we observe that at very small separations 
$R\simeq 10^{-3}$, the doubling time has a tendency to a constant value 
$\langle T(R)\rangle \simeq 0.0016$. On these scales we are below the forcing
scale (see Figure\ref{fig1}), and the velocity field can be assumed
smooth. As a consequence of Lagrangian chaos we expect on these scales an
exponential amplification of separations \cite{CFPV91} at a rate given by
the Lagrangian Lyapunov exponent $\lambda $. The latter can be obtained as 
$\lambda =\lim_{R\to 0}\ln \rho /\langle T(R)\rangle $ \cite{ABCCV97} and
gives $\lambda \simeq 110$ (in adimensional units). The Lagrangian Lyapunov
exponent $\lambda $ is a small scale quantity (i.e. depends on the Reynolds
number of the simulation), and thus has to be compared with a small scale
characteristic time. One can estimate the smallest characteristic time 
$\tau_{min}$ by the minimum value of $(k^{3}E(k))^{-1}$. We obtain 
$\lambda \simeq 0.23\tau _{min}^{-1}$.

The comparison of Figures \ref{fig2} and \ref{fig5} shows the
advantages of fixed scale statistics with respect to fixed time statistics.
In particular, the dependence of $R(t)$ and the Richardson constant on the
initial separation are absent in fixed scale statistics $T(R)$.

\begin{figure}[tbp]
\epsfbox{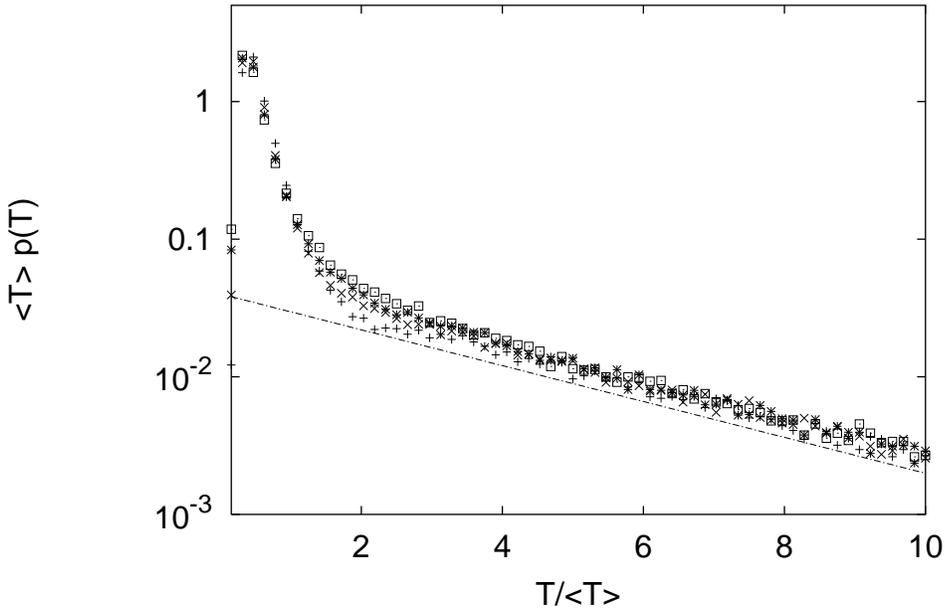}
\caption{ Pdf of doubling times for distances $R=0.003$ ($+$), $R=0.075$ 
($\times$), $R=0.02$ ($\ast$) and $R=0.046$ ($\Box$). The dashed line is the
exponential $exp(-0.3 \, T/\langle T \rangle)$. }
\label{fig6}
\end{figure}

In Figure~\ref{fig6} we plot the doubling time pdf $p_{D}(T)$
compensated with the mean value $\langle T(R)\rangle $ at different scales
in the inertial range $0.003\le R\le 0.046$. First, we obtain a very nice
collapse of the different curves, indicating that the process is really
self-similar. Second, we observe the exponential tail predicted in 
Section~\ref{sec:3} with a fitted coefficient $0.3$ which is indeed not 
far from the theoretical prediction (\ref{eq:3.5}) based on the 
Richardson's picture.
The difference between the predicted and measured values of the prefactors
is not large, but perceptible: it shows that the Richardson's equation
gives a correct qualitative description of the dispersion process, but is
not exact. The reasons for deviations from the diffusive picture proposed by
Richardson are the long-range correlations in the particles' motion, as seen
from the analysis of the turning points of their relative trajectories.

\subsection{Turning points statistics and the persistence parameter}
\label{sec:4.3} 

A possible explanation for the deviations of our high resolution numerical
data from the Richardson's picture is the not too small value of the
persistence parameter. As discussed in Section~\ref{sec:2}, at large values
of $Ps$ the contribution of ballistic events may lead to non-Richardson
distributions and moreover makes the dispersion strongly sensible to
finite-size effects, cutting the longer trajectories.

\begin{figure}[tbp]
\epsfbox{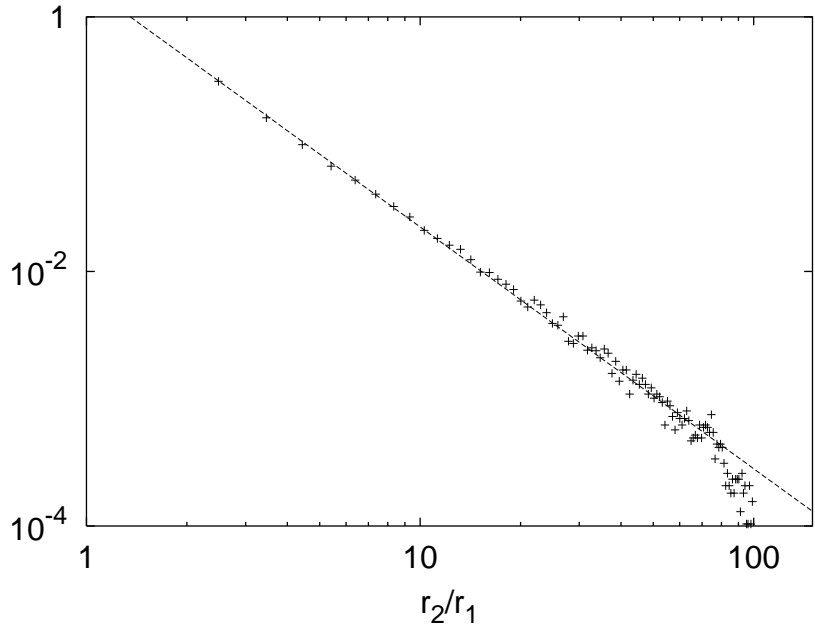}
\caption{ Probability density function of turning point ratio $\Psi(r_2/r_1)$.
The exponent of the power law (dashed line) gives the value $Ps \simeq 0.87$.}
\label{fig7}
\end{figure}

We have computed the persistence parameter making use of (\ref{eq:2.8}). We
have recorded, for each pair, the set of turning points $r_{i}$ at which the
pair's relative velocity changes sign. From the set of $r_{i}$ we have then
computed the pdf of the ratio $r_{i+1}/r_{i}$, accumulating for all the $i$
and all the pairs. 
The result, plotted in Figure~\ref{fig7}, gives $Ps\simeq 0.87$. 
The requirement that both $r_1$ and $r_2$ are in the inertial range,
strongly limits the statistics on turning points and the numerical result is 
affected by rather large uncertainty. Nevertheless, 
it is remarkable that the power law
tail in the conditional probability density $\Psi (r_{2}|r_{1})$ is well
observed in our numerical simulations. This justifies, a posteriori, the use
of models based on $\Psi (r_{2}|r_{1})$ for describing relative dispersion 
\cite{S99}. The numerical value of the effective persistence parameter 
$Ps\simeq 0.87$ is not so small, and can explain the observed deviations from
Richardson pdf (which are however less pronounced in a 2d flow than in a
theoretical one-dimensional model \cite{SKB00}): The transport in a 2d
turbulent flow is neither purely diffusive nor ballistic.

In order to be more confident on the numerical value of $Ps$ obtained 
through the turning-points statistics, let us show that it
agrees with a simple estimates based on the values of the
Kolmogorov's and the Richardson's constants. According to the Kolmogorov's
scaling, the mean squared relative velocity of the pair is given by 
\begin{equation}
\left\langle \delta v^{2}(r)\right\rangle =C_2\varepsilon ^{2/3}r^{2/3},
\label{eq:4.3}
\end{equation}
with $C_2 \simeq 12.9$ \cite{BCV00}.
If the particles separate ballistically with the rms velocity 
\begin{equation}
\delta v(r)=C_2^{1/2} \varepsilon ^{1/3}r^{1/3}
\label{eq:4.4}
\end{equation}
the distance between them should grow as 
\begin{equation}
R_{\max }^{2}=\left( \frac{2}{3}\right)^{3} C_2^{3/2} \varepsilon t^{3} \, .
\label{eq:4.5}
\end{equation}
On the other hand, due to the unsteadiness of the separation velocity, the
distance between the particles grows slower, namely as $R^{2}=g\varepsilon
t^{3}$, so that the factor 
\begin{equation}
\xi ^{2}=R^{2}/R_{\max }^{2}=\left(\frac{3}{2}\right)^{3}
\frac{g}{C_2^{3/2}}
\label{eq:4.6}
\end{equation}
serves as a measure of this unsteadiness and $\xi ^{2}$ is connected with 
the value of the persistence parameter. In our case, $\xi^{2}\approx 0.28$.
Within the stochastic model of Ref.~\cite{SKB00} this 
corresponds to a value of $Ps$ between $1.1$ and $1.2$, 
in reasonable agreement with the direct measurement
from the turning-point statistics, and again corroborates the stochastic
approach. 

\begin{figure}[tbp]
\epsfbox{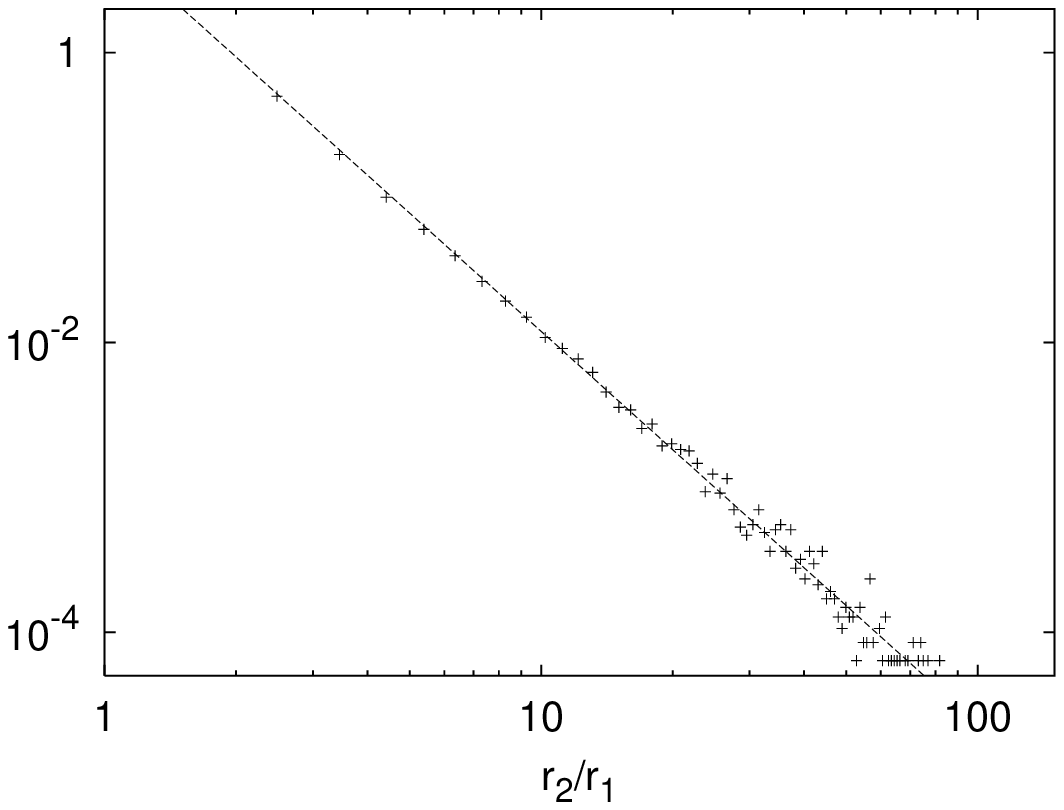}
\caption{ The same of Figure~\ref{fig7} but for $\lambda=0.5$. The
persistence parameter is now $Ps \simeq 0.58$.}
\label{fig8}
\end{figure}

We also note a possibility to ``tune'' the $Ps$ value by performing
simulations in which the Lagrangian trajectories are integrated according to 
$\dot{{\bf x}}=\lambda {\bf v}({\bf x},t)$. By changing the value of
parameter $\lambda $ one effectively changes $v_{0}$ and thus $Ps$. In the
extreme case $\lambda \to 0$ the trajectories resemble those in a time $%
\delta $-correlated velocity field. In the opposite limit, $\lambda >>1$ we
have dispersion on a quenched field. Of course, it is only for the standard
value $\lambda =1$ that Lagrangian trajectories move consistently with
velocity field (i.e. for $\nu =\alpha =\phi =0$ (\ref{eq:4.1}) conserves
vorticity along the Lagrangian trajectories). For other values of $\lambda$
such simulations suffer the typical problem of advection in synthetic field
(i.e. wrong reproduction of the sweeping effect, see \cite{BCCV99} for a
discussion). Simulations for several values of $\lambda $ show that
existence of the power-law tails of $\Psi (r_{2}|r_{1})$ is a robust effect,
as supposed by the model of Refs.\cite{S99,SKB00}, and that $Ps$ grows with 
$\lambda$. As an example, in Figure~\ref{fig8} we plot the probability
density $\Psi (r_{2}|r_{1})$ obtained from a simulation with $\lambda =0.5$.
All the Eulerian parameters are the same of Figure~\ref{fig7}. We again
observe a clear power law tail but now with $Ps\simeq 0.58$.


\section{Conclusions}

\label{sec:5}

In this paper we have investigated the Lagrangian relative dispersion in
direct numerical simulation of two-dimensional turbulence. The inverse
energy cascade of two-dimensional turbulence displays Kolmogorov scaling
without intermittency and it is thus the natural framework for investigating
possible deviations from the classical Richardson picture. Moreover, the
large enough values of the accessible Reynolds numbers make it possible to
obtain the quantitative results not inferior to those of laboratory
experiments.

The analysis of the numerical data was performed by using both standard
statistics at fixed time and exit time statistics at fixed scale. The latter
is shown to be more robust to finite Reynolds effects. An application of
exit time statistics is developed for measuring the Richardson constant. The
value found is somewhat larger than one obtained experimentally and in
kinematic simulations, and thus nearer to the predictions of the closure
theories.

The results of numerical simulation of statistics of the interparticle
distances are in good agreement with the original Richardson's description
in terms of diffusion equation. The large deviations (with respect to
Richardson theory) observed in the pdf of separation at fixed times are
mostly related to crossover effects due to finite Reynolds numbers and
disappear when looking at exit time statistics. The deviations found here
are only of the quantitative nature. Thus, the Richardson's equation gives a
good basis for qualitative description of the dispersion in turbulent flows.

The deviations found in the fixed-scale statistics and also the strong
sensitivity to the finite size effects can be associated with the
long-range-correlated nature of the particles' relative motion in turbulent
flow. Paying attention to the turning points of the relative trajectories
allows for estimating the effective persistence parameter of the motion
which is found to be of the order of unity. Thus, the motion shows a strong
ballistic component and is not purely diffusive. On the other hand, the
correlations are not too strong to fully destroy the Richardson's picture.

We note that the methodology of analysis proposed here based on the
fixed-scale statistics and on the analysis of the relative trajectories can
be also applied to the analysis of laboratory experiments. It would be
extremely interesting to see, to what extent the laboratory flows reproduce
the results reported here.

\begin{acknowledgments}
We gratefully acknowledge the support of the DFG through SFB428, of the
Fonds der Chemischen Industrie and of INFM (PRA-TURBO).
We thank M.~Cencini and J.~Klafter for useful discussions.
Numerical simulations were partially performed at CINECA within 
the INFM project ``Fully developed two-dimensional turbulence''.
\end{acknowledgments}

\appendix

\section{The PDF of Doubling Times}

\label{app:1}

Let us discuss the probability $p_{D}\left( t\right) $ of the time when the
a pair of particles initially at distance $R/\rho $ separates up to the
distance $R$, and obtain its asymptotic decay of this probability for $t$
large.

Changing to a variable $\xi =(k_{0}\varepsilon ^{1/3})^{-1/2}r^{1/3}$
reduces (\ref{eq:2.2}) to a radial part of a spherically symmetric
diffusion equation in $3d$ dimensions with constant diffusion coefficient.
In 2d one has 
\begin{equation}
\frac{\partial p}{\partial t}=\frac{1}{9\xi ^{5}}\frac{\partial }{\partial
\xi }\xi ^{5}\frac{\partial }{\partial \xi }p  
\label{eq:app1.1}
\end{equation}
with the initial condition $p(\xi ,0)=\delta (\xi _{\min }-\xi )$ with 
$\xi_{\min }=(k_{0}\varepsilon ^{1/3})^{-1/2}(R/\rho )^{1/3}$ and with the
boundary condition $p(\xi _{\max },t)=0$, with $\xi _{\max
}=(k_{0}\varepsilon ^{1/3})^{-1/2}R^{1/3}$.

The solution of a boundary-value problem for (\ref{eq:app1.1}) can be
obtained by means of eigenfunction decomposition. Assuming the variable
separation we get the solution in the form 
$p(\xi,t)=\sum_{i}e^{-\lambda_{i}^{2}t}\psi _{i}(\xi )$, where 
$\psi_{i}(\xi )$ is an eigenfunction of the equation 
\begin{equation}
\frac{1}{9\xi^{5}}\frac{\partial }{\partial \xi }\xi ^{5}\frac{\partial }
{\partial \xi }\psi_{i}=-\lambda _{i}^{2}\psi _{i}
\label{eq:app1.2}
\end{equation}
satisfying the boundary condition $\psi (\xi _{\max })=0$. The corresponding
solution which is nonsingular in zero is $\psi_{i}=\xi^{-2}J_{2}(3
\lambda_{i}\xi )$ ($J_{2}$ is the Bessel function \cite{AbraSteg}). The fact
that $\psi$ vanishes at $\xi_{\max }$ gives $3 \lambda _{i}
\xi_{max}=j_{2,i} $, where $j_{2,i}$ is the $i$-the real zero of $J_{2}(x)$.
For example, the smallest eigenvalue is 
$\lambda_{1}^{2}=j_{2,1}^{2}/9\xi_{\max}^{2} \approx 2.93 k_{0} 
\varepsilon^{1/3} R^{-2/3}$. 
Thus, the long-time asymptotic of the doubling-time distribution
is $\exp (-2.93 k_{0} \varepsilon^{1/3} R^{-2/3}t)$.


\section{Average doubling time}

\label{app:2}

The mean doubling time can be obtained from a stationary solution of the
Richardson diffusion equation. Imagine that one particle per unit time is
introduced at $r=R/\rho $ and there are respectively a reflecting and
absorbing boundaries at $r=0$ and $r=R$. The stationary solution of (\ref
{eq:2.2}) in 2d with the appropriate boundary conditions and continuity are 
$R/\rho $ is 
\begin{equation}
p(r)=\left\{ 
\begin{array}{lll}
C\left[ \rho ^{4/3}-1\right]  & \mbox{for} & 0<r<R/\rho  \\ 
C\left[ \left( {\frac{r}{R}}\right) ^{-4/3}-1\right]  & \mbox{for} & R/\rho
<r<R
\end{array}
\right.
\label{eq:app2.1}
\end{equation}
The number of particle in $r<R$ is 
\begin{equation}
N=\int_{|{\bf r}|<R}p(r)d{\bf r}=2\pi \int_{0}^{R}rp(r)dr
\label{eq:app2.2}
\end{equation}
where $S_{d}$ is a surface of a unit sphere in $d$ dimensions. By using (\ref
{eq:app2.1}) one obtains 
\begin{equation}
N={2\pi }C(1-\rho ^{-2/3})R^{d}.
\label{eq:app2.3}
\end{equation}

The current at $r=R$, i.e. the number of particle exiting from the boundary 
$R$ per unit time, is given, as in (\ref{eq:3.3}), as 
\begin{equation}
J=-2\pi \varepsilon ^{1/3}k_{0}R^{7/3}\left. {\frac{\partial p(r)}{\partial r%
}}\right| _{r=R}={\frac{8\pi }{3}}C\varepsilon ^{1/3}k_{0}R^{4/3}
\label{eq:app2.4}
\end{equation}

The mean doubling time is the average time spent by a particle at $r<R$. It
is given by the ratio $N/J$ and thus 
\begin{equation}
\langle T_{\rho }(R)\rangle ={\frac{3}{4}}{\frac{\rho ^{2/3}-1}{\varepsilon
^{1/3}k_{0}\rho ^{2/3}}}\,R^{2/3}
\label{eq:app2.5}
\end{equation}
which is Eq.(\ref{eq:3.4}).



\begin{references}
\bibitem{Batchelor52}  G.K. Batchelor, ``Diffusion in a field of homogeneous
turbulence II: The relative motion of particles,'' Proc. Camb. Phil. Soc. 
{\bf 48}, 345 (1952).

\bibitem{MY75}  A. Monin and A. Yaglom, {\it Statistical Fluid Mechanics \/}
(MIT Press, Cambridge, Mass., 1975), Vol. 2.

\bibitem{Richardson26}  L.F. Richardson, ``Atmospheric diffusion shown on a
distance-neighbour graph,'' Proc. Roy. Soc. A {\bf 110}, 709 (1926).

\bibitem{ZB94}  N. Zovari and A. Babiano, ``Derivation of the relative
dispersion law in the inverse energy cascade of two-dimensional
turbulence,'' Physica D {\bf 76}, 318 (1994).

\bibitem{EM96}  F.W. Elliott, Jr. and A.J. Majda, ``Pair dispersion over an
inertial range spanning many decades,'' Phys. Fluids {\bf 8}, 1052 (1996).

\bibitem{FV98}  J.C.H. Fung and J.C. Vassilicos, ``Two-particle dispersion
in turbulentlike flows,'' Phys. Rev. E {\bf 57}, 1677 (1998).

\bibitem{JPT99}  M.C. Jullien, J. Paret and P. Tabeling, ``Richardson pair
dispersion in two-dimensional turbulence,'' Phys. Rev. Lett. {\bf 82}, 2872
(1999).

\bibitem{MOA99}  J. Mann, S. Ott and J.S. Andersen, ``Experimental Study of
Relative, Turbulent Diffusion,'' Ris\o National Laboratory technical report
R-1036(EN) (1999).

\bibitem{BCCV99}  G. Boffetta, A. Celani, A. Crisanti and A. Vulpiani,
``Pair dispersion in synthetic fully developed turbulence,'' Phys. Rev. E 
{\bf 60}, 6734 (1999).

\bibitem{BC00}  G. Boffetta and A. Celani, ``Pair dispersion in
turbulence,'' Physica A {\bf 280} 1 (2000).

\bibitem{ABCCV97}  V. Artale, G. Boffetta, A. Celani, M. Cencini and A.
Vulpiani, ``Dispersion of passive tracers in closed basins: Beyond the
diffusion coefficient,'' Phys. Fluids A {\bf 9}, 3162 (1997).

\bibitem{Okubo62}  A. Okubo, ``A review of theoretical models for turbulent
diffusion in the sea,'' J. Oceanol. Soc. Japan {\bf 20}, 286 (1962).

\bibitem{K66} R. Kraichnan, ``Dispersion of particle pairs in homogeneous
turbulence,'' Phys. Fluids {\bf 9}, 1728 (1966).

\bibitem{S99} I.M. Sokolov, ``Two-particle dispersion by correlated random
velocity fields,'' Phys. Rev. E {\bf 60}, 5528 (1999).

\bibitem{K68} R. Kraichnan, ``Small-Scale Structure of a Scalar Field
Convected by Turbulence,'' Phys. Fluids {\bf 11}, 945 (1968).

\bibitem{GV00} K. Gawedzki and M. Vergassola, ``Phase transition in the
passive scalar advection,'' Physica D {\bf 138}, 63 (2000).

\bibitem{SKB00}  I.M. Sokolov, J. Klafter and A. Blumen, ``Ballistic versus
diffusive pair dispersion in the Richardson regime,'' Phys. Rev. E {\bf 61},
2717 (2000).

\bibitem{SWK87} M.F. Shlesinger, B. West and J. Klafter,
``L\'{e}vy dynamics of enhanced diffusion: Application to turbulence,''
Phys. Rev. Lett. {\bf 58}, 1100 (1987).

\bibitem{KM80}  R.H. Kraichnan and D. Montgomery, ``Two-dimensional
turbulence,'' Rep. Prog. Phys. {\bf 43}, 547 (1980).

\bibitem{BCV00}  G. Boffetta, A. Celani and M. Vergassola ``Inverse energy
cascade in two-dimensional turbulence: Deviations from Gaussian behavior,''
Phys. Rev. E {\bf 61}, R29 (2000)

\bibitem{SY93}  L.M. Smith and V. Yakhot, ``Bose Condensation and
Small-Scale Structure Generation in a Random Force Driven 2D Turbulence,''
Phys. Rev. Lett. {\bf 71}, 352 (1993).

\bibitem{FHMP92}  J. Fung, J. Hunt, N. Malick and R. Perkins ``Kinematic
simulation of homogeneous turbulence by unsteady random Fourier modes,'' J.
Fluid Mech. {\bf 236}, 281 (1992).

\bibitem{LL81}  M.~Larcheveque and M.~Lesieur, ``The application of
eddy-dumped Markovian closures to the problem of dispersion of particle
pairs,'' J. Mech. {\bf 20}, 113 (1981).

\bibitem{CFPV91}  A. Crisanti, M. Falcioni, G. Paladin and A. Vulpiani,
``Lagrangian Chaos: Transport, Mixing and Diffusion in Fluids,'' Riv. Nuovo
Cimento {\bf 14}, 1 (1991).

\bibitem{AbraSteg}  M. Abramovitz and I.A. Stegun, eds., {\it Handbook of
mathematical functions}, Dover, N.Y. 1972
\end{references}
\end{document}